\documentstyle[12pt,fleqn]{article}

\font\twlgot =eufm10 scaled \magstep1
\font\egtgot =eufm8
\font\sevgot =eufm7
\font\twlmsb =msbm10 scaled \magstep1
\font\egtmsb =msbm8
\font\sevmsb =msbm7

\newfam\gotfam

\textfont\gotfam\twlgot
\scriptfont\gotfam\egtgot
\scriptscriptfont\gotfam\sevgot

\newfam\msbfam
\textfont\msbfam\twlmsb
\scriptfont\msbfam\egtmsb
\scriptscriptfont\msbfam\sevmsb

\def\pBbb{\relax\ifmmode\expandafter\Bb\else\typeout{You cann't use
Bbb in text mode}\fi}
\def\Bb #1{{\fam\msbfam\relax#1}}

\def\thebibliography#1{\section*{References}\list
    {[\arabic{enumi}]}{\settowidth\labelwidth{#1}\leftmargin\labelwidth
      \advance\leftmargin\labelsep
      \usecounter{enumi}}
      \def\newblock{\hskip .11em plus .33em minus .07em}
      \sloppy\clubpenalty4000\widowpenalty4000
      \sfcode`\.=1000\relax}

\let\Large=\large

\def\op#1{\mathop{\fam0 #1}\limits}

\newcommand{\id}{{\rm Id\,}}

\newcommand{\beq}{\begin{equation}}
\newcommand{\eeq}{\end{equation}}
\newcommand{\ben}{\begin{eqnarray}}
\newcommand{\een}{\end{eqnarray}}
\newcommand{\be}{\begin{eqnarray*}}
\newcommand{\ee}{\end{eqnarray*}}
\newcommand{\bea}{\begin{eqalph}}
\newcommand{\eea}{\end{eqalph}}
\newcommand{\cA}{{\cal A}}

\newcommand{\cD}{{\cal D}}

\newcommand{\cH}{{\cal H}}

\newcommand{\bL}{{\bf L}}
\newcommand{\bR}{{\bf R}}
\newcommand{\bC}{{\bf C}}
\newcommand{\bZ}{{\bf Z}}

\newcommand{\bt}{\beta}

\newcommand{\la}{\lambda}
\newcommand{\La}{\Lambda}
\newcommand{\f}{\phi}

\newcommand{\Om}{\Omega}
\newcommand{\m}{\mu}

\newcommand{\vt}{\vartheta}

\newcommand{\lng}{\langle}
\newcommand{\rng}{\rangle}

\newcommand{\si}{\sigma}

\newcommand{\w}{\wedge}

\newcommand{\wh}{\widehat}
\newcommand{\ol}{\overline}
\newcommand{\dr}{\partial}

\newcommand{\ot}{\otimes}

\let\ssection=\section
\renewcommand{\section}{\setcounter{equation}{0}\ssection}

\newcounter{eqalph}
\newcounter{equationa}
\newcounter{remark}
\newcounter{example}
\newcounter{theorem}
\newcounter{proposition}
\newcounter{lemma}
\newcounter{corollary}
\newcounter{definition}
\newenvironment{eqalph}{\stepcounter{equation}
\setcounter{equationa}{\value{equation}}
\setcounter{equation}{0}

\begin{eqnarray}}{\end{eqnarray}\setcounter{equation}{\value{equationa}}}

\def\theremark{\arabic{remark}}
\def\therexample{\arabic{remark}}

\def\thedefinition{\arabic{definition}}

\newcommand{\mar}[1]{}

\begin{document}
\hbox{}

{\parindent=0pt

{\Large \sc Geometric quantization of completely integrable Hamiltonian systems
in the action-angle variables}

\bigskip

{\sc G.Giachetta}$^a$,\footnote{{\it E-mail addresses}:
giovanni.giachetta@unicam.it (G.Giachetta), luigi.mangiarotti@unicam.it
(L.Mangiarotti),
sard@grav.phys.msu.su (G.Sardanashvily)}
{\sc L.Mangiarotti}$^a$ and {\sc G.Sardanashvily}$^b$
\bigskip

$^a$ {\it Department of Mathematics and Physics, University of Camerino,
62032 Camerino (MC), Italy}

$^b$ {\it Department of Theoretical Physics,
Moscow State University, 117234 Moscow, Russia}
\bigskip

{\small
{\bf Abstract.}
\medskip
}

We provide geometric quantization of a completely integrable
Hamiltonian system in the action-angle
variables around an invariant torus with respect to polarization
spanned by almost-Hamiltonian vector fields of angle
variables. The associated quantum algebra consists
of functions affine in action coordinates. We obtain a set
of its nonequivalent representations in
the separable pre-Hilbert space of smooth complex
functions on the torus where action operators and a Hamiltonian
are diagonal and have countable spectra.

\medskip


}

\bigskip
\bigskip

We study geometric quantization of
an autonomous completely integrable system
around a regular compact invariant
manifold. By virtue of the classical Arnold--Liouville theorem, a small
neighbourhood $W$ of this manifold in the ambient momentum phase space
$M$ is isomorphic to the symplectic annulus
\mar{00}\beq
W=V\times T^m, \label{00}
\eeq
where $V\subset \bR^m$
is a nonempty domain
and $T^m$ is an $m$-dimensional torus [1,2].
The product (\ref{00}) is equipped with
the action-angle coordinates $(I_k,\f^k)$.
With respect to these coordinates, the
symplectic form on $W$ (\ref{00}) reads
\mar{ci1}\beq
\Om=dI_k\w d\f^k, \label{ci1}
\eeq
while a Hamiltonian of a completely
integrable system is a function $\cH(I_k)$ of action variables only.

There are different approaches to quantization of completely integrable
Hamiltonian systems [3,4]. The advantage of the geometric
quantization procedure
is that it remains equivalent under symplectic diffeomorphisms.
Geometric quantization of
completely integrable Hamiltonian systems has been studied
with respect to polarization spanned by Hamiltonian vector fields of
integrals of motion [5,6]. In fact, the Simms quantization of
the harmonic oscillator is also of this type [7].
The problem is that the associated quantum algebra
includes functions which are not globally defined,
and elements of the carrier space are not smooth
sections of a quantum bundle.  Indeed, written with respect to the
action-angle variables, this quantum
algebra consists of functions which are affine in angle coordinates.

Here, we use a different polarization. Since a Hamiltonian of a completely
integrable system on the symplectic annulus $W$ (\ref{00})
depends only on action variables, it seems natural
to provide the Schr\"odinger representation of action variables by first order
differential operators on functions of angle coordinates. For this purpose,
one should choose the angle polarization of the symplectic annulus (\ref{00})
spanned by almost-Hamiltonian vector fields $\dr^k$ of angle variables.

Since the action-angle coordinates are canonical for the symplectic
form (\ref{ci1}), geometric quantization
of the symplectic annulus $(W,\Om)$ in fact is equivalent to 
geometric quantization
of the cotangent bundle $T^*T^m=\bR^m\times T^m$ of the torus $T^m$
provided with the canonical symplectic form. In particular, the
above mentioned angle polarization
of $V\times T^m$ corresponds to the familiar vertical polarization of
$T^*T^m$ which leads to Schr\"odinger quantization of the cotangent
bundle $T^*T^m$. The associated quantum algebra $\cA$ of $W$
consists of functions which are affine in action variables
$I_k$. It possesses the continuum set of nonequivalent representations
by first order differential operators
in the separable pre-Hilbert space $\bC^\infty(T^m)$
of smooth complex functions
on $T^m$. This set is indexed by homomorphisms of the de Rham cohomology
group $H^1(T^m)=\bR^m$ of the torus $T^m$ to the circle group $U(1)$.

As is well known, an application of Schr\"odinger geometric quantization
is limited by the fact that a Hamiltonian fails to belong to the 
quantum algebra $\cA$,
unless it is affine in momenta. In the case of a completely integrable system
on the symplectic annulus $W$, a Hamiltonian $\cH$ depends only on 
action variables which
mutually commute. Therefore, if $\cH(I_k)$ is a polynomial function, 
it is uniquely
represented by an element of the enveloping algebra $\ol\cA$ of the Lie algebra
$\cA$, and is quantized as the operator $\cH(\wh I_k)$.
Moreover, this quantization is also extended to
Hamiltonians which are analytic functions on $\bR^m$ because, as we 
will observe, the
action operators $\wh I_k$ are diagonal.

As was mentioned above, the geometric quantization procedure
is equivalent under symplectic diffeomorphisms, but it essentially depends
on the choice of polarization [8,9]. Given a symplectic 
diffeomorphism (\ref{00})
of $W$ to the product $V\times T^m$, geometric quantization of $V\times T^m$
with respect to the angle polarization of $V\times T^m$
implies the equivalent quantization of the
initial completely integrable system on a neighbourhood $W$ of a 
compact invariant
manifold with respect to the induced angle polarization of $W$. However, this
polarization of $W$ is not canonical because an isomorphism (\ref{00})
by no means is unique.
Furthermore, there are topological obstructions
to global action-angle coordinates [10,11]. Therefore, quantization with
respect to the angle polarization is not extended to
the whole momentum phase space $M$ of a completely integrable system. 
For instance,
one usually mentions a harmonic oscillator as the simplest completely
integrable Hamiltonian system whose quantization in the action-angle variables
looks notoriously difficult because the eigenvalues of its action operator
is expected to be lower bounded (see [12] for a survey). However, a
harmonic oscillator written
relative to action-angle coordinates $(I,\f)$ is located in
the momentum phase space
$\bR^2\setminus\{0\}$, but it is not the standard oscillator on
$\bR^2$. Namely, there is a monomorphism, but not an isomorphism of the
Poisson algebra of smooth complex functions on $\bR^2$ to that on
$\bR^2\setminus\{0\}$. Furthermore,
the angle polarization on $\bR^2\setminus\{0\}$
is not extended to $\bR^2$. As a consequence, the quantum algebra associated
to this polarization contains functions on $\bR^2\setminus\{0\}$ which
are not extended to $\bR^2$, and its carrier
space $\bC^\infty(T^m)$ does so. As was mentioned above, the Simms
quantization of the
harmonic oscillator on the momentum phase space
$\bR^2\setminus\{0\}$ with respect to the polarization spanned by the
Hamiltonian vector field of a Hamiltonian $\cH$ is
quantization with respect to the action-angle variables.
The carrier space of this quantization consists of tempered distributions,
and the spectrum of the Hamiltonian is lower bounded [7].

In accordance with the standard geometric quantization procedure [13,14],
since the symplectic form $\Om$ (\ref{ci1}) is exact, the prequantum bundle
is defined as a trivial complex line bundle $C$ over $V\times T^m$.
Since the action-angle coordinates are canonical for the symplectic
form (\ref{ci1}), the prequantum bundle $C$ need no metaplectic
correction, and it is a
quantum bundle.
Let its trivialization
\mar{ci3}\beq
C\cong (V\times T^m) \times \bC \label{ci3}
\eeq
hold fixed. Any other trivialization leads to
equivalent quantization of $V\times T^m$.
Given the associated bundle coordinates $(I_k,\f^k,c)$, $c\in\bC$, on
$C$ (\ref{ci3}),
one can treat its sections as smooth complex functions on
$V\times T^m$.

The Konstant--Souriau prequantization formula associates to
each smooth real function $f\in C^\infty(V\times T^m)$ on
$V\times T^m$ the first order differential operator
\mar{lqq46}\beq
\wh f=-i\nabla_{\vt_f} + f \label{lqq46}
\eeq
on sections of $C$, where
\be
\vt_f=\dr^kf\dr_k -\dr_kf\dr^k
\ee
is the Hamiltonian vector field of $f$ and
$\nabla$ is the covariant differential with respect to a
suitable $U(1)$-principal connection on $C$. This connection
preserves the
Hermitian metric $g(c,c')=c\ol c'$ on $C$, and
its curvature form obeys the prequantization
condition $R=i\Om$. It reads
\mar{ci20}\beq
A=A_0 +icI_kd\f^k\ot\dr_c \label{ci20}
\eeq
where $A_0$ is a flat $U(1)$-principal connection on $C\to
V\times T^m$. The classes of gauge nonconjugated flat
principal connections on $C$ are indexed
by homomorphisms of the de Rham cohomology
group
\be
H^1(V\times T^m)=H^1(T^m)=\bR^m
\ee
of the annulus $V\times T^m$ to $U(1)$ [7], i.e.,
their set is bijective to $\bR^m/\bZ^m$.
We choose their representatives of the form
\be
A_0[(\la_k)]=dI_k\ot\dr^k + d\f^k\ot(\dr_k +i\la_kc\dr_c),
\qquad \la_k\in [0,1).
\ee
Accordingly, the relevant connection (\ref{ci20}) on $C$ up to
gauge conjugation reads
\mar{ci14}\beq
A[(\la_k)]=dI_k\ot\dr^k + d\f^k\ot(\dr_k +i(I_k+\la_k)c\dr_c).
   \label{ci14}
\eeq
For the sake of simplicity, we will assume that the numbers $\la_k$
in the expression(\ref{ci14}) belongs to $\bR$, but will bear in mind that
connections $A[(\la_k)]$ and $A[(\la'_k)]$ with $\la_k-\la'_k\in\bZ$
are gauge conjugated. Given a connection (\ref{ci14}),
the prequantization operators (\ref{lqq46}) read
\mar{ci4}\beq
\wh f=-i\vt_f +(f-(I_k+\la_k)\dr^kf). \label{ci4}
\eeq

Let us choose the above mentioned angle polarization $V\pi$ which is
the vertical tangent bundle of the fibration $\pi:V\times T^m\to T^m$.
This polarization is spanned by the vectors $\dr^k$.
It is readily observed that the corresponding quantum algebra
$\cA\subset C^\infty(V\times T^m)$
consists of affine functions
\mar{ci13}\beq
f=a^k(\f^r)I_k + b(\f^r) \label{ci13}
\eeq
of action coordinates $I_k$.
The carrier space of its representation by operators (\ref{ci4}) is
defined as the
space $E$ of sections $\rho$ of the prequantum bundle $C$ of
compact support
which obey the condition $\nabla_\vt\rho=0$ for any Hamiltonian vector field
$\vt$ subordinate to the distribution $V\pi$. This condition reads
\be
\dr_kf\dr^k\rho=0, \qquad \forall f\in C^\infty(T^m).
\ee
It follows that elements of $E$ are independent of action variables and,
consequently, fail to be of compact support, unless $\rho=0$.
This is the well-known problem of Schr\"odinger geometric quantization.
It is solved as follows [15,16].

Given an imbedding $i_T:T^m\to V\times T^m$,
let $C_T=i^*_TC$ be the pull-back of the prequantum bundle $C$ (\ref{ci3})
over the torus $T^m$. It is a trivial complex line bundle $C_T=T^m\times\bC$
provided with the pull-back Hermitian metric
$g(c,c')=c\ol c'$. Its sections are smooth complex functions on
$T^m$. Let
\be
\ol A = i^*_TA= d\f^k\ot(\dr_k +i(I_k+\la_k)c\dr_c)
\ee
be the pull-back of the connection $A$ (\ref{ci14}) onto $C_T$.
Let $\cD$ be a metalinear bundle
of complex half-forms on the torus $T^m$.
It admits the canonical lift
of any vector field $\tau$ on $T^m$, and
the corresponding
Lie derivative of its sections reads
\be
\bL_\tau=\tau^k\dr_k+\frac12\dr_k\tau^k.
\ee
Let us consider the tensor product
\mar{ci6}\beq
Y=C_T\ot\cD\to T^m. \label{ci6}
\eeq
Since the Hamiltonian vector fields
\be
\vt_f=a^k\dr_k-(I_r\dr_ka^r +\dr_kb)\dr^k
\ee
of functions $f$ (\ref{ci13}) are projectable onto $T^m$, one can
associate to each
element $f$ of the quantum algebra $\cA$ the first order
differential operator
\mar{lmp135}\beq
\wh f=(-i\ol\nabla_{\pi\vt_f} +f)\ot\id+\id\ot\bL_{\pi \vt_f}=
-ia^k\dr_k-\frac{i}{2}\dr_ka^k-a^k\la_k +b \label{lmp135}
\eeq
on sections of $Y$. A direct
computation shows
that the operators (\ref{lmp135}) obey the Dirac condition
\be
[\wh f,\wh f']=-i\wh{\{f,f'\}}.
\ee
Sections $s$ of the quantum bundle $Y\to T^m$ (\ref{ci6})
constitute a pre-Hilbert space $E_T$ with respect to the nondegenerate
Hermitian
form
\be
\lng s|s'\rng=\left(\frac1{2\pi}\right)^m\op\int_{T^m}
s \ol s, \qquad s,s\in E_T.
\ee
Then it is readily observed that $\wh f$ (\ref{lmp135}) are Hermitian operators
in $E_T$. They provide a desired Schr\"odinger geometric quantization of
a completely integrable Hamiltonian system on the annulus $V\times T^m$.
Of course, this quantization depends on the choice of a
connection $A[(\la_k)]$ (\ref{ci14}) and
a metalinear bundle $\cD$. The latter need not be trivial.

If
$\cD$ is trivial, sections of the quantum bundle $Y\to T^m$ (\ref{ci6})
obey the transformation
rule
\be
s(\f^k+2\pi)=s(\f^k)
\ee
for all indices $k$. They are naturally complex smooth functions on $T^m$.
By virtue of the multidimensional Fourier theorem [17], the functions
\mar{ci15}\beq
\psi_{(n_r)}=\exp[i(n_r\f^r)], \qquad (n_r)=(n_1,\ldots,n_m)\in\bZ^m,
\label{ci15}
\eeq
constitute an orthonormal basis for the pre-Hilbert space
$E_T=\bC^\infty(T^m)$.
The action operators
\mar{ci7}\beq
\wh I_k=-i\dr_k -\la_k \label{ci7}
\eeq
(\ref{lmp135}) are diagonal
\mar{ci9}\beq
\wh I_k\psi_{(n_r)}=(n_k-\la_k)\psi_{(n_r)} \label{ci9}
\eeq
with respect to this basis. Other elements of the algebra $\cA$ are decomposed
into the pull-back functions $\pi^*\psi_{(n_r)}$
which act on $\bC^\infty(T^m)$ by multiplications
\mar{ci11}\beq
\pi^*\psi_{(n_r)} \psi_{(n'_r)}=\psi_{(n_r)}
\psi_{(n'_r)}=\psi_{(n_r+n'_r)}. \label{ci11}
\eeq

If $\cD$ is a nontrivial metalinear bundle, sections of the quantum bundle
$Y\to T^m$ (\ref{ci6}) obey the transformation
rule
\mar{ci8}\beq
\rho_T(\f^j+2\pi)=-\rho_T(\f^j)  \label{ci8}
\eeq
for some indices $j$. In this case, the orthonormal basis of the
pre-Hilbert space
$E_T$ can be represented by double-valued complex functions
\mar{ci10}\beq
\psi_{(n_i,n_j)}=\exp[i(n_i\f^i+ (n_j+\frac12)\f^j)] \label{ci10}
\eeq
on $T^m$. They are eigenvectors
\be
\wh I_i\psi_{(n_i,n_j)}=(n_i-\la_i)\psi_{(n_i,n_j)}, \qquad
\wh I_j\psi_{(n_i,n_j)}=(n_j-\la_j +\frac12)\psi_{(n_i,n_j)}
\ee
of the operators $\wh I_k$ (\ref{ci7}), and the pull-back functions
$\pi^*\psi_{(n_r)}$
act on the basis (\ref{ci10}) by the above law (\ref{ci11}).
It follows that the representation of the quantum algebra $\cA$
determined by the connection
$A[(\la_k)]$ (\ref{ci14}) in the space of sections
(\ref{ci8}) of a nontrivial quantum bundle $Y$ (\ref{ci6})
is equivalent to its representation determined by the connection
$A[(\la_i,\la_j-\frac12)]$ in the space $\bC^\infty(T^m)$
of smooth
complex functions on $T^m$.

Therefore, one can restrict the study of representations of
the quantum algebra $\cA$ to its representations in $\bC^\infty(T^m)$
associated
to different connections (\ref{ci14}). These representations are
nonequivalent, unless
   $\la_k-\la'_k\in\bZ$ for all indices $k$.

Given the representation (\ref{lmp135}) of the quantum algebra $\cA$
in $\bC^\infty(T^m)$,
any polynomial Hamiltonian $\cH(I_k)$ of
a completely integrable system is uniquely quantized as a Hermitian element
$\wh\cH(I_k)=\cH(\wh I_k)$ of the enveloping algebra
$\ol\cA$ of $\cA$. It has the countable spectrum
\be
\wh \cH(I_k)\psi_{(n_r)}=\cH(n_k-\la_k) \psi_{(n_r)}.
\ee
Note that, since $\wh I_k$ are diagonal, one can also quantize Hamiltonians
$\cH(I_k)$ which are analytic functions on $\bR^m$.

As a conclusion remark, let us assume that
a Hamiltonian
$\cH$ of a completely integrable system is independent of action variables
$I_a$ ($a,b,c=1,\ldots,l$). Then its eigenvalues are countably
degenerate. Let us consider the perturbed Hamiltonian
\mar{01}\beq
\cH'=\Delta(\xi^\m,\f^b,I_a) +\cH(I_j), \label{01}
\eeq
where the perturbation term $\Delta$ depends on the action-angle
coordinates with the above mentioned indices
$a,b,c,\ldots$ and on some time-dependent parameters $\xi^\la(t)$ by the law
\mar{02}\beq
\Delta=\La^a_\la(\xi^\m,\f^b)\dr_t\xi^\la I_a. \label{02}
\eeq
The Hamiltonian $\cH'$ characterizes a Hamiltonian system with time-dependent
parameters [18-21]. Being affine in action variables, the perturbation
term $\Delta$ (\ref{02}) admits the instantwise quantization
  by the operator
\be
\wh\Delta=-(i\La^a_\bt\dr_a +\frac{i}{2}\dr_a\La^a_\bt
  +\la_a \La^a_\bt)\dr_t\xi^\bt.
\ee
Since the operators $\wh\Delta$ and $\wh\cH$ mutually commute, the
total quantum evolution operator reduces to the product
\be
T\exp\left[-i\op\int_0^t\wh\cH'dt'\right]=
U_1(t)\circ U_2(t)=
T\exp\left[-i\op\int_0^t\wh\cH dt'\right]\circ
T\exp\left[-i\op\int_0^t\wh\Delta dt'\right].
\ee
The first factor $U_1$ in this product is the dynamic evolution operator of
the quantum completely integrable Hamiltonian system.
The second operator acts in the eigenspaces of the operator $U_1$
and reads
\mar{zz25}\ben
&& U_2(t)=T\exp\left[\op\int_0^t\{
-\La^a_\bt(\f^b,\xi^\m(t'))\dr_a -\frac12\dr_a\La^a_\bt(\f^b,\xi^\m(t'))
+i\la_a \La^a_\bt(\f^b,\xi^\m(t'))\}\dr_t\xi^\bt dt'\right]\nonumber\\
&&\qquad =T\exp\left[\op\int_{\xi([0,t])}\{
-\La^a_\bt(\f^b,\si^\m)\dr_a -\frac12\dr_a\La^a_\bt(\f^b,\si^\m)
+i\la_a \La^a_\bt(\f^b,\si^\m)\} d\si^\bt\right]. \label{zz25}
\een
It is readily observed that this operator depends on the curve
$\xi([0,1])\subset S$ in the parameter space $S$. One can treat it as
an operator of parallel
displacement along the curve
$\xi$ [19-22].
For instance, if $\xi([0,1])$ is a loop in $S$, the operator $U_2$
(\ref{zz25}) is
the geometric Berry factor. In this case, one can think of $U_2$ as
being a holonomy control operator [23].
At present, such control operators
attract special attention in connection with
holonomic quantum computation [24-26].

\end{document}